\documentstyle[12pt]{article}
\begin{document}

\begin{center}

{\sf \Large Modified 2D Proca theory: off-shell nilpotent symmetries and their 
mathematical as well as physical implications}

\vskip 2.5 cm

{\sf{R. P. Malik$^{(a,b)}$}}\\
{\it $^{(a)}$Physics Department, BHU-Varanasi-221 005, India}\\
{\it $^{(b)}$DST-CIMS, Faculty of Science, BHU-Varanasi-221 005, India}\\
{\small {\sf {E-mails: rpmalik1995@gmail.com; rudra.prakash@hotmail.com}}}
\end{center}

\vskip 2.5cm

\noindent
{\sf Abstract:} We derive the off-shell nilpotent Becchi-Rouet-Stora-Tyutin (BRST),
anti-BRST, (anti-)co-BRST, a bosonic and  the ghost-scale symmetry transformations for a couple 
of {\it equivalent} Lagrangian densities of the two (1 + 1)-dimensional (2D)
Stueckelberg-modified version of Proca theory which {\it also} incorporates a pseudo-scalar field. 
We also discuss algebraically suitable discrete symmetry transformations of the theory. Finally, we demonstrate the relevance of the above continuous and discrete
symmetries in the context of  differential geometry and establish that our present 
{\it massive} 2D theory
is a tractable field theoretic model for the Hodge theory. One of the key novel observations
of our present investigation is the appearance of Curci-Ferrari (CF) type restrictions 
{\it even} in the case of our present {\it massive} 2D Abelian 1-form gauge theory. 
We also point out the mathematical as well as physical implications of the 
above pseudo-scalar field in our present theory.\\



\newpage

\section {Introduction}

The Proca theory, in any arbitrary dimension of spacetime, 
is a generalized version of the Maxwell's 
theory which describes a {\it massive} bosonic field. In physical 
four (3 + 1)-dimensions of spacetime,
the above bosonic field is endowed with three degrees of 
freedom due to its mass. The beautiful
gauge symmetry of the Maxwell's theory is, however, {\it not} respected by the 
Proca theory because the latter
is endowed with second-class constraints in the language of Dirac's prescription for classification
scheme (see, e.g. [1,2]).  It is the Stueckelberg's formalism (see, e.g. [3])
that converts the above second-class
constraints into first-class constraints thereby restoring the original gauge symmetry of the
Maxwell's theory in the case of Proca theoy which is modified by the inclusion of a real scalar field.

In our earlier works [4-8] on the $p$-form (with $p = 1, 2, 3$) Abelian gauge theories
in $D = 2 \,p$ dimensions of spacetime, we have provided physical realizations of the 
de Rham cvohomological operators\footnote{On a compact manifold without
a boundary, a set of three mathematical operators $(d, \delta, \Delta)$ are called the
de Rham cohomological operators of differential geometry where $d = dx^\mu \partial_\mu$ 
(with $d^2 = 0$) is the exterior derivative, $\delta = \pm * d *$ (with $\delta^2 = 0$)
is the co-exterior derivatives and $\Delta = (d + \delta)^2 \equiv \{d, \delta \}$ is the
Laplacian operator. Here the $*$ symbol stands for
 the Hodge duality operation on the compact
manifold. The above operators obey a well-known cohomological algebra [cf. equation (37) below].}
and Hodge duality operation of differential geometry 
within the framework of BRST formalism where the continuous and discrete symmetry transformations
have played a key role. These theories are, however, {\it not} massive gauge theories
as there is no mass parameter in them. In a very recent set of papers [9-11], we
have been able to prove that the ${\cal N} = 2$ supersymmetric quantum mechanical models 
are {\it also} the examples of Hodge theory because of their specific discrete
and continuous symmetries. Our present 2D modified Proca theory is very {\it special} 
because it is  a {\it massive} gauge theory.

One of the central themes of our present investigation is to focus on the Stueckelberg-modified
version of Proca theory in two (1 + 1)-dimensions of spacetime and demonstrate that this theory
is a model for the Hodge theory where {\it mass} and various kinds of 
{\it internal} symmetries co-exist together in a meaningful manner if we incorporate a pseudo-scalar field in the theory with appropriate mass dimension (in natural units). 
The inclusion
of the latter field enhances the symmetry properties of the theory and it adds aesthetics
to the mathematical as well as physical contents of this 2D massive Abelian gauge theory.

Our present kind of studies is physically as well as mathematically useful because, exploiting
the key mathematical 
ideas and theoretical tools of such studies, we have been able to prove that the 2D
(non-)Abelian 1-form gauge theories (without any interaction with matter fields)
are a new class of topological field theories (TFTs) which capture in their folds a few aspects
of Witten type TFT and some salient features of Schwarz type TFT 
(see, e.g. [4,12,13] for details). In this proof, the idea of Hodge decomposition
theorem in the quantum Hilbert space of states has played a key role [14].
Furthermore, we have also been able to show that 
the 2D Abelian
theory with Dirac fields is a model for Hodge theory where a TFT (i.e. the 2D Abelian 1-form gauge field) couples with the matter 
(Dirac) fields [15,16]. Such studies have 
also been helpful in showing that the free 4D Abelian 2-form and 6D Abelian 3-form gauge 
theories are models for the quasi-TFT [5-8].

In our present investigation, we have come across some {\it novel} features while proving 
our modified model of 2D Proca theory to be an example of Hodge theory
where the nilpotent (co-)exterior derivatives of differential geometry have been identified
with the off-shell nilpotent (anti-)co-BRST and (anti-)BRST symmetry transformations. 
 For instance, we have found the existence of a set of two equivalent Lagrangian densities
for the description of this 2D theory where a {\it new} pseudo-scalar field has been introduced
on mathematical and physical grounds. Further, we have found that, even for this Abelian 1-form
gauge theory, there is existence of CF-type restriction [17]
when we discuss the symmetry properties
of the {\it equivalent} Lagrangian densities. Finally, the pseudo-scalar field appears
in the theory with negative kinetic term and, hence, could be a candidate for the dark matter.  
The above features have, hitherto, not been seen {\it together} in the context of 
BRST approach to the description of $D = 2p$ dimensional 
Abelian $p$-form ($ p = 1, 2, 3)$ gauge theories [4-8].

The main motivating factors behind our present investigation are as 
follows. 

First and foremost, we have to generalize our earlier work [18] on 
modified version  of Proca theory where we have
discussed the on-shell nilpotent (anti-)BRST, (anti-)co-BRST, a unique bosonic and the ghost-scale
symmetries.

Second, it is well-known that the off-shell nilpotent symmetries are more general
than the on-shell nilpotent symmetries as the latter are the special case of the former. 
We have accomplished this goal in our present investigation.

Third, our present endeavor is our modest step towards our main goal of finding the physically interesting
4D {\it massive} models for the Hodge theories where {\it mass} and various 
kinds of internal symmetries would co-exist together in a meaningful manner.

Finally,
we have to explore the possibility of the existence of CF-type restriction(s) which
are the hallmarks (see, e.g.  [19,20]) of a given gauge theory described within the framework 
of BRST formalism. In fact, we do find something like CF-type restrictions when we discuss
the symmetry properties of the two equivalent Lagrangian densities. However, in terms of properties, the CF-type restriction  of our present theory is somewhat different from the usual
CF-condition found in the context of 4D non-Abelian 1-form gauge theory.

The contents of our present investigation are organized as follows. To set up the notations, we recapitulate
the bare essentials of the (dual-)gauge symmetries for the gauge-fixed Lagrangian 
densities of the Proca theory
and discuss its generalizations. We also discuss the discrete symmetries of this generalized version of the
Lagrangian densities of Proca theory in Sec. 2. Our Sec. 3 is devoted to the derivation of the off-shell
nilpotent (anti-)BRST symmetry transformations. In our Sec. 4, 
we deduce the off-shell nilpotent (anti-)co-BRST
symmetry transformations. Our Sec. 5 contains
 a unique bosonic symmetry of our present theory. In Sec. 6,
we discuss the ghost-scale symmetry and full discrete symmetries of our present theory. Our Sec. 7 presents
a thorough discussion of the extended BRST algebra and its connections with the cohomological operators. Finally,
we make some concluding remarks and point out a few future directions in our Sec. 8.

{\it General notations and conventions:}
Through out the whole body of our text, we shall denote the (anti-)BRST, (anti-)co-BRST, a bosonic and the ghost-scale symmetry transformations
 by the notations $s_{(a)b}, s_{(a)d}, s_w$ and
$s_g$, respectively. For the two equivalent Lagrangian densities of our present theory, 
we shall use the superscripts $(1, 2)$ to distinguish them from each-other.
We shall confine ourselves to only {\it internal} symmetry transformations and we shall  
{\it not} discuss anything connected with the 2D spacetime symmetries  of our present theory as
the spacetime 2D Minkowski manifold always remains in the background and it does not 
actively participate
in our whole discussion.

\section {Local (dual-)gauge transformations}

We begin with the modified 
version of the two (1 + 1)-dimensional (2D) Proca theory which is described
by the recently proposed {\it equivalent} gauge-fixed Lagrangian densities
${\cal L}_{(1,2)}$ as given below\footnote{We adopt 
here the convention and notations such that the flat background
2D Minkowskian spacetime manifold is endowed with a metric $\eta_{\mu\nu}$ with
signatures $(+1, -1)$ so that $A\cdot B = \eta_{\mu\nu} A^\mu B^\nu \equiv A_0 B_0 - A_1 B_1$
is the dot product between two non-bull vectors $A_\mu$ and $B_\mu$. 
Here the Greek indices $\mu, \nu, \lambda, ... = 0, 1$ denote the time and
space directions of the 2D spacetime manifold. We also use the 2D antisymmetric Levi-Civita tensor
$\varepsilon_{\mu\nu}$ with the convention $\varepsilon_{01} = + 1 = \varepsilon^{10}$
and $\varepsilon^{\mu\nu}\, \varepsilon_{\mu\nu} = - 2!, \; \varepsilon^{\mu\nu}\, \varepsilon_{\mu\lambda} = - 1! \delta^\nu_\lambda, \;    \varepsilon^{\mu\nu} \, 
\varepsilon_{\nu\lambda} = \delta^\mu_\lambda$, etc.}
(see, e.g. [18] for details)
\begin{eqnarray} 
{\cal L}_{(1,2)} &=& \frac {1}{2}\, {(E \mp m\,\tilde\phi)}^2 \pm m\, E\,\tilde\phi - \frac {1}{2}\,\partial_\mu \,\tilde\phi\,\,\partial^\mu\,\tilde\phi 
+ \frac {m^2}{2} A_\mu\, A^ \mu + \frac {1}{2}\, \partial_\mu\, \phi\, \partial^\mu\, \phi 
\nonumber\\ &\mp& m \,A_\mu \,\partial^\mu\, \phi
 -\frac {1}{2}\,(\partial\cdot A \pm m\,\phi)^2,
\end{eqnarray} 
where the electric field $E = - \varepsilon^{\mu\nu} \partial_\mu A_\nu \equiv F_{01}$ is the only
existing component of the 2D curvature tensor $F_{\mu\nu} = \partial_\mu A_\nu - \partial_\nu A_\mu$
which is derived from the curvature 2-form $F^{(2)} = d A^{(1)} = [(dx^\mu \wedge dx^\nu)/2!]\, F_{\mu\nu}$.
Here $d = dx^\mu \partial_\mu$ (with $d^2 = 0$) is the exterior derivative and 1-form $A^{(1)} = dx^\mu A_\mu$
defines the Abelian 1-form connection field $A_\mu$.  In exactly similar fashion, the gauge-fixing term
$(\partial \cdot A)$ owes its origin to the co-exterior derivative $\delta = - \,* \,d \,*$ (with $\delta^2 = 0$)
because $\delta A^{(1)} = (\partial \cdot A)$ where $*$ is the Hodge 
duality operation on the 2D flat
Minkowskian spacetime manifold. There is {\it no} magnetic field in the 2D theory.

In the above Lagarangian densities, the basic fields ($A_\mu, \phi, \tilde \phi$) have mass dimension zero in
natural units (i.e. $\hbar = c = 1$) and the parameter $m$ has the dimension of mass. Thus, the latter is the
mass parameter in the whole theory. The pair ($\phi, \tilde \phi$) are the real-scalar and pseudo-scalar fields
as are the gauge-fixing term $(\partial \cdot A)$ and the electric field component $E$, respectively. On
mathematical grounds, we are free to add/subtract the pair of fields $(\phi, \tilde \phi)$ to the gauge-fixing
and kinetic terms with proper mass dimension. This is what has been precisely achieved  in (1). In fact, the signatures
of these fields, play very
important roles in the discussion of discrete symmetries as would become clear later.
It is worth pointing out that the origin of the field $\phi$ lies in 
the Stueckelberg formalism which is normally adopted
to restore the gauge symmetry in a massive gauge theory (see, e.g. [3]).

The modification of the gauge-fixing term and kinetic term has been done with proper care where the signatures
of addition/subtraction of the pair of fields ($\phi, \tilde \phi)$ play very crucial roles. It can be seen
that, under the following (dual-)gauge transformations $\delta_{(d)g}$
\begin{eqnarray}
&& \delta_{dg} A_\mu = - \varepsilon_{\mu\nu} \partial^\nu \Sigma, \qquad \delta_{dg} E = \Box \Sigma, 
\qquad \delta_{dg} \tilde \phi = \mp\,m\,\Sigma, \nonumber\\
&& \delta_{dg} (\partial \cdot A) = 0, \qquad \delta_{dg} \phi = 0, \qquad 
\delta_{dg} (\partial \cdot A + m \phi) = 0,
\end{eqnarray}
 \begin{eqnarray}
&& \delta_{g} A_\mu = \partial_\mu \Lambda, \qquad \delta_{g} E = 0, 
\qquad  \delta_{g} (E - m \tilde \phi) = 0, \nonumber\\
&& \delta_{g} (\partial \cdot A) =  \Box \Lambda, \qquad \delta_{g} \phi = \pm \,m \, \Lambda, \qquad 
\delta_{g} \tilde \phi = 0,
\end{eqnarray}
the Lagrangian densities ${\cal L}_{(1, 2)}$ transform as follows:
\begin{eqnarray}
\delta_{dg} {\cal L}_{(1, 2)} &=& \partial_\mu \Bigl [ m \,\varepsilon^{\mu\nu}\; 
\bigl (m \,A_\nu \,\Sigma
\pm \,\phi\, \partial_\nu \,\Sigma \bigr ) \pm m \,\tilde \phi \,\partial^\mu \,\Sigma \Bigr ]
+ \,(E \mp m \,\tilde \phi) \, (\Box + m^2) \, \Sigma, \nonumber\\
\delta_{g} {\cal L}_{(1, 2)} &=& - (\partial \cdot A\,  \pm \,m \, \phi) \, (\Box + m^2) \, \Lambda.
\end{eqnarray}
Thus, we note that, to attain the {\it perfect} (dual-)gauge symmetry, we have to impose exactly similar
kind of restrictions on the (dual-)gauge transformation parameters $(\Sigma)\Lambda$. In other
words, we have to put the mathematical conditions:  $(\Box + m^2) \, \Sigma = 0,
(\Box + m^2) \, \Lambda = 0$ on the local (dual-)gauge
parameters $(\Sigma) \Lambda$ where $\Box = \partial_0^2 - \partial_1^2$.

There is a perfect discrete symmetry in the theory, even though, 
there is modification of the kinetic
and gauge-fixing terms due to inclusion of the fields $\tilde \phi$ and $\phi$. For instance, it 
can be readily checked that under the following discrete symmetry transformations
\begin{eqnarray}
A_\mu \to \pm i \varepsilon_{\mu\nu} A^\nu, \quad \phi \to \pm i \tilde \phi, 
\quad \tilde \phi \to \pm i  \phi, \quad E \to \mp i (\partial \cdot A), \quad
(\partial \cdot A) \to \mp i E,
\end{eqnarray} 
the Lagrangian densities ${\cal L}_{(1,2)}$ remain invariant modulo some total spacetime derivatives.
We would like to mention, in passing, that the above discrete symmetries (and their generalizations) 
would play very important roles in our subsequent discussions.

The kinetic and gauge-fixing terms can be linearized by invoking the Nakanishi-Lautrup type
auxiliary fields (${\cal B}, \bar {\cal B}, B, \bar B $). The linearized versions of the above Lagrangian densities ${\cal L}_{(1,2)}$ are different in appearance and these are as follows:
\begin{eqnarray} 
{\cal L}_{(b_1)} &=& {\cal B} \, {(E - m\,\tilde\phi)}^2  - \frac{1}{2} \, {\cal B}^2
+ m\, E\,\tilde\phi - \frac {1}{2}\,\partial_\mu \,\tilde\phi\,\,\partial^\mu\,\tilde\phi 
+ \frac {m^2}{2} A_\mu\, A^ \mu + \frac {1}{2}\, \partial_\mu\, \phi\, \partial^\mu\, \phi 
\nonumber\\ &-& m \, A_\mu \,\partial^\mu\, \phi
 + B\,(\partial\cdot A +  m\,\phi) + \frac{1}{2}\, B^2,
\end{eqnarray} 
\begin{eqnarray} 
{\cal L}_{(b_2)} &=& \bar {\cal B} \, {(E + m\,\tilde\phi)}^2  - \frac{1}{2} \, \bar {\cal B}^2
- m\, E\,\tilde\phi - \frac {1}{2}\,\partial_\mu \,\tilde\phi\,\,\partial^\mu\,\tilde\phi 
+ \frac {m^2}{2} A_\mu\, A^ \mu + \frac {1}{2}\, \partial_\mu\, \phi\, \partial^\mu\, \phi 
\nonumber\\ &+& m A_\mu \,\partial^\mu\, \phi
 + \bar B\,(\partial\cdot A -  m\,\phi) + \frac{1}{2}\, \bar B^2.
\end{eqnarray} 
It is worthwhile to explain that the
Lagrangian densities ${\cal L}_{(1,2)}$  have been written in 
independent forms mainly due to the different types of Nakanishi-Lautrup auxiliary fields 
(${\cal B}, \bar {\cal B}, B, \bar B$)
that have been invoked, in the above, for the linearization purposes.

It is straightforward to check that there is a perfect discrete symmetry in the theory because the
following transformations 
\begin{eqnarray}
&& A_\mu \to \pm i \varepsilon_{\mu\nu} A^\nu, \quad \phi \to \pm i \tilde \phi, 
\quad \tilde \phi \to \pm i  \phi, \quad E \to \mp i (\partial \cdot A), \quad
(\partial \cdot A) \to \mp i E, \nonumber\\
&& B \to \pm\, i\, {\cal B}, \qquad \bar B \to \pm\, i\, \bar {\cal B},
\qquad {\cal  B} \to \pm\, i\, B, \qquad \bar {\cal  B} \to \pm\, i\, \bar B,
\end{eqnarray}   
leave the above Lagrangian densities  ${\cal L}_{(b_1, b_2)}$ invariant. We point out that 
the above transformations are nothing but the generalization of our 
earlier transformations (5).
In exactly, similar fashion, it can be seen that, under the following generalized 
local, continuous and infinitesimal (dual-)gauge
transformations $\delta^{(1)}_{(D)G}$:
\begin{eqnarray}
&& \delta^{(1)}_{DG}\, A_\mu = - \varepsilon_{\mu\nu} \partial^\nu \Sigma, \qquad 
\delta^{(1)}_{DG} \,E = \Box \Sigma, 
\qquad \delta^{(1)}_{DG} \,\tilde \phi = - \,m\,\Sigma, \qquad 
\delta^{(1)}_{DG} \,B = \delta^{(1)}_{DG} \,{\cal B} = 0, \nonumber\\
&& \delta^{(1)}_{DG} \,(\partial \cdot A) = 0, \quad \delta^{(1)}_{DG} \,\phi = 0, \quad 
\delta^{(1)}_{DG} \,(\partial \cdot A + m \phi) = 0, \quad 
\delta^{(1)}_{DG} \,(E - m \tilde \phi) = (\Box + m^2) \, \Sigma,
\nonumber\\
&& \delta^{(1)}_{G}\, A_\mu = \partial_\mu \Lambda, \qquad \delta^{(1)}_{G} \,E = 0, 
\qquad  \delta^{(1)}_{G} \,(E - m \tilde \phi) = 0, \qquad 
\delta^{(1)}_{G} \,B = \delta^{(1)}_{G} \,{\cal B} = 0,\nonumber\\
&& \delta^{(1)}_{G}\, (\partial \cdot A) =  \Box \Lambda, \quad 
\delta^{(1)}_{G} \,\phi = + \,m \, \Lambda, \quad 
\delta^{(1)}_{G} \,\tilde \phi = 0,
 \quad \delta^{(1)}_G \,(\partial \cdot A + m \phi) = (\Box + m^2) \, \Lambda,
\end{eqnarray}
the Lagrangian density ${\cal L}_{(b_1)}$ transforms as follows
\begin{eqnarray}
\delta^{(1)}_{DG} {\cal L}_{(b_1)} &=& \partial_\mu \,\Bigl [ m \,\varepsilon^{\mu\nu} \bigl (m \,A_\nu\, \Sigma
+ \phi \,\partial_\nu \,\Sigma \bigr ) + m \,\tilde \phi \,\partial^\mu \,\Sigma \Bigr ]
+ {\cal B} \, (\Box + m^2) \, \Sigma, \nonumber\\
\delta^{(1)}_{G} {\cal L}_{(b_1)} &=& B\;  (\Box + m^2) \, \Lambda.
\end{eqnarray}
which shows that, to achieve the (dual-)gauge invaraince in the theory, the (dual-)gauge parameters 
have to restricted in exactly similar fashion [i.e. $ (\Box + m^2) \, \Sigma = 0, \Box + m^2) \, \Lambda = 0$].

We can discuss the local, continuous and infinitesimal (dual-)gauge
transformations for the Lagrangian density ${\cal L}_{(b_2)}$, too. These transformations are
\begin{eqnarray}
&& \delta^{(2)}_{DG} \, A_\mu = - \varepsilon_{\mu\nu} \partial^\nu \Sigma, 
\qquad \delta^{(2)}_{DG} \,E = \Box \Sigma, 
\qquad \delta^{(2)}_{DG} \,\tilde \phi = + \,m\,\Sigma, 
\qquad \delta^{(2)}_{DG} \,\bar B = \delta^{(2)}_{DG} \,\bar {\cal B} = 0, \nonumber\\
&& \delta^{(2)}_{DG} \,(\partial \cdot A) = 0, \quad \delta^{(2)}_{DG} \,\phi = 0, \quad 
\delta^{(2)}_{DG} \,(\partial \cdot A - m \phi) = 0, \quad 
\delta^{(2)}_{DG}\, (E + m \tilde \phi) = (\Box + m^2) \, \Sigma,
\nonumber\\
&& \delta^{(2)}_{G} \, A_\mu = \partial_\mu \Lambda, \qquad \delta_{G}\, E = 0, 
\qquad  \delta^{(2)}_{G} \, (E + m \tilde \phi) = 0, \qquad 
\delta^{(2)}_{G} \, \bar B = \delta^{(2)}_{G} \,\bar {\cal B} = 0,\nonumber\\
&& \delta^{(2)}_{G} \,(\partial \cdot A) =  \Box \Lambda, \quad 
\delta^{(2)}_{G} \,\phi = - \,m \, \Lambda, \quad 
\delta^{(2)}_{G} \,\tilde \phi = 0 \quad 
\delta^{(2)}_G \,(\partial \cdot A - m \phi) = (\Box + m^2) \, \Lambda,
\end{eqnarray}
under which, as expected, the Lagrangian density ${\cal L}_{(b_2)}$ transforms as:
\begin{eqnarray}
\delta^{(2)}_{DG} {\cal L}_{(b_2)} &=& \partial_\mu \,\Bigl [ m \,\varepsilon^{\mu\nu} \bigl (m \,A_\nu\, \Sigma
- \phi \,\partial_\nu \,\Sigma \bigr ) - m \,\tilde \phi \,\partial^\mu \,\Sigma \Bigr ]
+ \bar {\cal B} \, (\Box + m^2) \, \Sigma, \nonumber\\
\delta^{(2)}_{G} {\cal L}_{(b_2)} &=& \bar B\;  (\Box + m^2) \, \Lambda.
\end{eqnarray}
Thus, once again, it is clear that for the sake of maintaining the (dual-)gauge symmetry 
invariance in the theory,
exactly similar kind of restrictions should be imposed on the infinitesimal (dual-)gauge parameters
$(\Sigma)\Lambda$ [i.e. $ (\Box + m^2) \, \Sigma = 0, \Box + m^2) \, \Lambda = 0$].

We wrap up this section with the following comments. First, the 
nomenclature of the (dual-)gauge transformations 
is very appropriate because we observe that, under the gauge transformations, it is the total kinetic
term, owing its fundamental origin to the exterior derivative, remains invariant. On the other hand,
it is the total guage-fixing term, originating 
basically from the dual-exterior derivative, that remains
unchanged under the dual-gauge transformations. Second, the above restrictions, on the (dual-)gauge 
transformation parameters, can be taken care of within the framework of BRST formalism where these
parameters would be replaced by 
the (anti-)ghost fields and there would be existence of {\it perfect} 
continuous as well as discrete symmetries in the BRST-invariant theory. Finally, we obtain the
following equations of motion from the Lagrangian densities ${\cal L}_{(b_1, b_2)}$:
\begin{eqnarray}
{\cal B} = E - m\, \tilde \phi, \quad  
\bar {\cal B} = E + m\, \tilde \phi,   \quad
B = \,- \,[(\partial \cdot A) - m\, \phi],  \quad 
\bar B = \,- \,[(\partial \cdot A) + m\, \phi],
\end{eqnarray}
which show that the Nakanishi-Lautrup type auxiliary fields are connected to one-another, via
some relevant fields of our present theory, in the following fashion:
\begin{eqnarray}
&& B + \bar B + 2 (\partial \cdot A) = 0, \qquad B - \bar B + 2\, m\, \phi = 0, \nonumber\\
&& {\cal B} + \bar {\cal B} - 2 \, E = 0, \qquad \qquad {\cal B} 
- \bar {\cal B} + 2\, m\, \tilde \phi = 0.
\end{eqnarray}
It is interesting to note that, under the {\it perfect} discrete symmetry transformations, these
relations transform amongst themselves. These relations are like the celebrated Curci-Ferrari (CF)
condition [17] of the non-Abelian 1-form gauge theories which appear in the BRST description
of the latter theories. We shall see, in our later discussions, that the above relations
do capture some properties of the CF-condition but there are  distinct differences, too. 
We shall dwell on this issue in our forthcoming sections in an explicit manner.

\section{Nilpotent (anti-)BRST symmetries}

The Lagrangian density ${\cal L}_{(b_1)}$ [cf. (6)]
can be generalized to the (anti-)BRST invariant
Lagrangian density ${\cal L}_B$
that incorporates the gauge-fixing and Faddeev-Popov ghost terms. 
The exact expression for this Lagrangian density is: 
\begin{eqnarray} 
{\cal L}_{(B)} &=& {\cal B} \, {(E - m\,\tilde\phi)}  - \frac{1}{2} \, {\cal B}^2
+ m\, E\,\tilde\phi - \frac {1}{2}\,\partial_\mu \,\tilde\phi\,\,\partial^\mu\,\tilde\phi 
+ \frac {m^2}{2} A_\mu\, A^ \mu + \frac {1}{2}\, \partial_\mu\, \phi\, \partial^\mu\, \phi 
\nonumber\\ &-& m \, A_\mu \,\partial^\mu\, \phi
 + B\,(\partial\cdot A +  m\,\phi) + \frac{1}{2}\, B^2 - i\,\partial_\mu\, \bar C \,
\partial^\mu \,C + i \,m^2 \,\bar C\, C,
\end{eqnarray} 
where the (anti-)ghost fields $(\bar C )C$ are fermionic (i.e. $C^2 = \bar C^2 = 0, C \bar C + \bar C C = 0$)
in  nature and they are required in the theory for the validity of unitarity. We observe that, under
the following (anti-)BRST transformations $s^{(1)}_{(a)b}$:
\begin{eqnarray}
&& s^{(1)}_{ab} \,A_\mu = \partial_\mu \bar C, \qquad 
s^{(1)}_{ab} E = s^{(1)}_{ab} \tilde \phi = s^{(1)}_{ab} \bar C = 0, 
\qquad  
s^{(1)}_{ab} B = s^{(1)}_{ab} {\cal B} = 0,\nonumber\\
&&  s^{(1)}_{ab} \phi = + \,m \, \bar C, \qquad  s^{(1)}_{ab}  C = -\, i\, B, \qquad 
s^{(1)}_{ab} \,(\partial \cdot A + m \phi) = (\Box + m^2) \, \bar C, \nonumber\\
&& s^{(1)}_{b} \,A_\mu = \partial_\mu  C, \qquad 
s^{(1)}_{b} E = s^{(1)}_{b} \tilde \phi = s^{(1)}_{b} C = 0, 
\qquad  
s^{(1)}_{b} B = s^{(1)}_{b} {\cal B} = 0,\nonumber\\
&&  s^{(1)}_{b} \phi = + \,m \,  C, \qquad  s^{(1)}_{b} \bar  C = +\, i\, B, \qquad 
s^{(1)}_{b} \,(\partial \cdot A + m \phi) = (\Box + m^2) \, C, 
\end{eqnarray}
the above Lagrangian density transforms to the total spacetime derivatives:
\begin{eqnarray}
s^{(1)}_b \,{\cal L}_{(B)} = \partial_\mu \,\Bigl [ \,B \,\partial^\mu \,C \, \Bigr ], 
\qquad  \qquad
s^{(1)}_{ab} \,{\cal L}_{(B)} = \partial_\mu \,\Bigl [ \,B \,\partial^\mu \, \bar C \, \Bigr ].
\end{eqnarray}
As a consequence, the action integral $S = \int dx\, {\cal L}_{(B)}$ remains invariant.

Exactly in a similar fashion, the Lagrangian density ${\cal L}_{(b_2)}$ can be generalized to
an equivalent (i.e. equivalent to  ${\cal L}_{(B)}$)
(anti-)BRST invariant Lagrangian density ${\cal L}_{(\bar B)}$:
\begin{eqnarray} 
{\cal L}_{(\bar B)} &=& \bar {\cal B} \, {(E + m\,\tilde\phi)}  - \frac{1}{2} \, \bar {\cal B}^2
- m\, E\,\tilde\phi - \frac {1}{2}\,\partial_\mu \,\tilde\phi\,\,\partial^\mu\,\tilde\phi 
+ \frac {m^2}{2} A_\mu\, A^ \mu + \frac {1}{2}\, \partial_\mu\, \phi\, \partial^\mu\, \phi 
\nonumber\\ &+& m \, A_\mu \,\partial^\mu\, \phi
 + \bar B\,(\partial\cdot A -  m\,\phi) + \frac{1}{2}\, \bar B^2 
- i\,\partial_\mu\, \bar C \,\partial^\mu \,C + i\, m^2 \,\bar C\, C,
\end{eqnarray}  
which also incorporates the gauge-fixing and Faddeev-Popov ghost terms. 
Furthermore, it is worth pointing
out that the ghost part of the Lagrangian densities ${\cal L}_{(B, \bar B)}$ is 
exactly the same [cf. (15), (18)]. 
The following (anti-)BRST symmetry transformations:
\begin{eqnarray}
&& s^{(2)}_{ab} \,A_\mu = \partial_\mu \bar C, \qquad 
s^{(2)}_{ab} E = s^{(2)}_{ab} \tilde \phi = s^{(2)}_{ab} \bar C = 0, \qquad  
s^{(2)}_{ab} \bar B = s^{(2)}_{ab} \bar {\cal B} = 0,\nonumber\\
&&  s^{(2)}_{ab} \phi = - \,m \, \bar C, \qquad  s^{(2)}_{ab}  C = -\, i\, \bar B, \qquad 
s^{(2)}_{ab} \,(\partial \cdot A - m \phi) = (\Box + m^2) \, \bar C, \nonumber\\
&& s^{(2)}_{b} \,A_\mu = \partial_\mu  C, \qquad 
s^{(2)}_{b} E = s^{(2)}_{b} \tilde \phi = s^{(2)}_{b} C = 0, 
\qquad  
s^{(2)}_{b} \bar B = s^{(2)}_{b} \bar {\cal B} = 0,\nonumber\\
&&  s^{(2)}_{b} \phi = - \,m \,  C, \qquad  s^{(2)}_{b} \bar  C = +\, i\, \bar B, \qquad 
s^{(2)}_{b} \,(\partial \cdot A - m \phi) = (\Box + m^2) \, C, 
\end{eqnarray}
leave the action integral invariant because the Lagrangian density  ${\cal L}_{(\bar B)}$
transforms to the total spacetime derivatives under the above symmetry transfromations:
\begin{eqnarray}
s^{(2)}_b \,{\cal L}_{(\bar B)} = \partial_\mu \,\Bigl [ \,\bar B \,\partial^\mu \,C \,\Bigr ],  \qquad \qquad
s^{(2)}_{ab} \,{\cal L}_{(\bar B)} = \partial_\mu \,\Bigl [ \,\bar B \,\partial^\mu \,
\bar C \,\Bigr ].
\end{eqnarray}
The above observation establishes that the (anti-)BRST transformations $s^{(2)}_{(a)b}$
are the {\it symmetry} transformations for the Lagrangian density  ${\cal L}_{(\bar B)}$. 

The noteworthy points, at this juncture, are as follows. First, the (anti-)BRST symmetry transformations
are off-shell  nilpotent [i.e. $ (s^{(1,2)}_{(a)b})^2 = 0$] of 
order two and they are absolutely
anticommuting (i.e. $s^{(1)}_b s^{(1)}_{ab} 
+ s^{(1)}_{ab} s^{(1)}_b = 0, s^{(2)}_b s^{(2)}_{ab} 
+ s^{(2)}_{ab} s^{(2)}_b = 0$) which demonstrates their fermionic nature and linear independence. Second,
the total kinetic term, owing its true origin to the exterior derivative of differential geometry, remains
invariant under the (anti-)BRST symmetry transformations. Finally, the absolute anticommutativity property of
the (anti-)BRST symmetry transformations, however, physically imply that only {\it one} of them could be identified
with the exterior derivative of differential geometry.

We close this section with the discussion on the  transformation 
properties of the Lagrangian density
${\cal L}_{(\bar B)}$  under the (anti-)BRST 
symmetry transformations $s^{(1)}_{(a)b}$. Furthermore, we
also devote time on the discussion of the transformation property of ${\cal L}_{(B)}$ 
under the (anti-)BRST symmetry
transformations $s^{(2)}_{(a)b}$
[cf. (19)]. We shall see that, in this exercise, one of the CF-type conditions [cf. (14)] 
would play an important role. 
Taking into account the following (anti-)BRST transformations, besides (16) and (19), namely;
\begin{eqnarray}
&& s^{(1)}_b \bar B = - 2\,\Box\,C, \qquad s^{(1)}_{ab} \bar B = - 2\,\Box\, \bar C, \qquad
s^{(1)}_b \bar {\cal B} = 0, \qquad s^{(1)}_{ab} \bar {\cal B} = 0, \nonumber\\
&& s^{(2)}_b  B = - 2\,\Box\,C, \qquad s^{(2)}_{ab}  B = - 2\,\Box\, \bar C, \qquad
s^{(2)}_b  {\cal B} = 0, \qquad s^{(2)}_{ab}  {\cal B} = 0,
\end{eqnarray}
we obtain the following expressions for the transformation properties of  ${\cal L}_{(\bar B)}$
and ${\cal L}_{(B)}$: 
\begin{eqnarray}
s^{(1)}_b {\cal L}_{(\bar B)} &=& \partial_\mu \, \Bigl [ B \,\partial^\mu \,C 
+ 2\, m^2\, A^\mu \,C + 2 \,m \,\phi \,\partial^\mu \,C \, \Bigr ] \nonumber\\
&-& \,\Bigl [ B + \bar B + 2 \,(\partial \cdot A) \Bigr ] \,( \Box + m^2) C, \nonumber\\
s^{(1)}_{ab} {\cal L}_{(\bar B)} &=& \partial_\mu \, \Bigl [ B \,\partial^\mu \,\bar C 
+ 2\, m^2\, A^\mu \,\bar C + 2 \,m \,\phi \,\partial^\mu \,\bar C \, \Bigr ] \nonumber\\
&-& \,\Bigl [ B + \bar B + 2 \,(\partial \cdot A) \Bigr ] \,( \Box + m^2) \bar C, \nonumber\\
s^{(2)}_{b} {\cal L}_{( B)} &=& \partial_\mu \, \Bigl [ \bar B \,\partial^\mu \, C 
+ 2 \,m^2\, A^\mu\,  C - 2\, m \,\phi \,\partial^\mu \, C \, \Bigr ] \nonumber\\
&-& \,\Bigl [ B + \bar B + 2 \,(\partial \cdot A) \Bigr ]\, ( \Box + m^2) C, \nonumber\\
s^{(2)}_{ab} {\cal L}_{(B)} &=& \partial_\mu \, \Bigl [ \bar B \,\partial^\mu \, \bar C 
+ 2 \,m^2\, A^\mu\, \bar  C - 2\, m \,\phi \,\partial^\mu \, \bar C \, \Bigr ] \nonumber\\
&-& \,\Bigl [ B + \bar B + 2 \,(\partial \cdot A) \Bigr ]\, ( \Box + m^2) \bar C,
\end{eqnarray}
which shows that $s^{(1, 2)}_{(a)b}$ are {\it also}  the symmetry transformations for the
Lagrangian densities ${\cal L}_{(\bar B)}$ and ${\cal L}_{( B)}$, respectively,
 provided we restrict ourselves to a 
constrained hypersurface 
in the 2D Minkowskian spacetime manifold where the (anti-)BRST invariant 
(i.e. $s_{(a)b} \, [B + \bar B + 2 (\partial \cdot A)] = 0$)  CF-type restriction 
$B + \bar B + 2 (\partial \cdot A) = 0$ [(cf. (14)] is satisfied. This observation is exactly like
the role played by the 
{\it original} CF-condition [17] in the BRST description of the 4D non-Abelian 
1-form gauge theory (see, e.g. [21,22] for more details). However, the CF-type condition,
present in our theory (i.e.  $B + \bar B + 2 (\partial \cdot A) = 0$), does {\it not} play any
role in the proof of  anticommutativity of the (anti-)BRST symmetry
transfromations of our present
theory as the latter are, as pointed out earlier, already absolutely anticommuting
(i.e. $s^{(1)}_b s^{(1)}_{ab} + s^{(1)}_{ab} s^{(1)}_b = 0, 
s^{(2)}_b s^{(2)}_{ab} + s^{(2)}_{ab} s^{(2)}_b = 0$) 
in nature. 

\section{Nilpotent (anti-)co-BRST symmetries}

Exactly  like the (anti-)BRST symmetry transformations, there are other fermionic symmetry
transformations
in the theory. For instance,  we observe that, under the following fermionic 
(anti-)dual-BRST [or (anti-)co-BRST] symmetry transformations $s^{(1)}_{(a)d}$:
\begin{eqnarray}
&& s^{(1)}_{ad} \,A_\mu = - \varepsilon_{\mu\nu} \partial^\nu  C, \qquad 
s^{(1)}_{ad} (\partial \cdot A) = s^{(1)}_{ad} \phi = s^{(1)}_{ad}  C = 0, 
\qquad  
s^{(1)}_{ad} B = s^{(1)}_{ad} {\cal B} = 0,\nonumber\\
&&  s^{(1)}_{ad} \tilde \phi = - \,m \,  C, \quad  s^{(1)}_{ad} \bar C = +\, i\, {\cal B}, \quad 
s^{(1)}_{ad} \,(E - m \tilde \phi) = (\Box + m^2) \,  C, \quad s^{(1)}_{ad} E = \Box C, \nonumber\\
&& s^{(1)}_{d} \,A_\mu = - \varepsilon_{\mu\nu} \partial^\nu  \bar C, \qquad 
s^{(1)}_{d} (\partial \cdot A) = s^{(1)}_{d} \phi = s^{(1)}_{d}  \bar C = 0, 
\qquad  
s^{(1)}_{d} B = s^{(1)}_{d} {\cal B} = 0,\nonumber\\
&&  s^{(1)}_{d} \tilde \phi = - \,m \,  \bar C, \quad  s^{(1)}_{d}  C = -\, i\, {\cal B}, \quad 
s^{(1)}_{d} \,(E - m \tilde \phi) = (\Box + m^2) \, \bar C, \quad s^{(1)}_{d} E = \Box \bar C,
\end{eqnarray}
the Lagrangian density ${\cal L}_{(B)}$ transforms to the total spacetime derivatives: 
\begin{eqnarray}
&&s^{(1)}_{ad} \,{\cal L}_{(B)} = \partial_\mu \,\Bigl [ \,{\cal B} \,\partial^\mu \,C 
+ m \, \varepsilon^{\mu\nu}\, \bigl (m \,A_\nu \,C + \phi\, \partial_\nu \,C \bigr )
+ m \,\tilde \phi \,\partial^\mu \,C \Bigr ], \nonumber\\ 
&& s^{(1)}_d \,{\cal L}_{(B)} = \partial_\mu \,\Bigl [ \,{\cal B}\, \partial^\mu \,\bar C 
+ m \, \varepsilon^{\mu\nu}\, \bigl (m\, A_\nu \,\bar C 
+ \phi \,\partial_\nu \,\bar C \bigr )
+ m \, \tilde \phi\, \partial^\mu\, \bar C \Bigr ],
\end{eqnarray}
which shows that the above transformations are the {\it symmetry} transformations for the action
integral $S = \int \, dx \, {\cal L}_{(B)}$ of our present theory.

We can discuss about the (anti-)co-BRST symmetry transformations for the equivalent Lagrangian density 
${\cal L}_{(\bar B)}$, too. These transformations are listed below: 
\begin{eqnarray}
&& s^{(2)}_{ad} \,A_\mu = - \varepsilon_{\mu\nu} \partial^\nu  C, \qquad 
s^{(2)}_{ad} (\partial \cdot A) = s^{(2)}_{ad} \phi = s^{(2)}_{ad}  C = 0, 
\qquad  
s^{(2)}_{ad} \bar B = s^{(2)}_{ad} \bar {\cal B} = 0,\nonumber\\
&&  s^{(2)}_{ad} \tilde \phi = + \,m \,  C, \quad  s^{(2)}_{ad} \bar C = +\, i\, \bar {\cal B}, \quad 
s^{(2)}_{ad} \,(E + m \tilde \phi) = (\Box + m^2) \,  C, \quad s^{(2)}_{ad} E = \Box C, \nonumber\\
&& s^{(2)}_{d} \,A_\mu = - \varepsilon_{\mu\nu} \partial^\nu  \bar C, \qquad 
s^{(2)}_{d} (\partial \cdot A) = s^{(2)}_{d} \phi = s^{(2)}_{d}  \bar C = 0, 
\qquad  
s^{(2)}_{d} \bar B = s^{(1)}_{d} \bar {\cal B} = 0,\nonumber\\
&&  s^{(2)}_{d} \tilde \phi = + \,m \,  \bar C, \quad  s^{(2)}_{d}  C = -\, i\, \bar {\cal B}, \quad 
s^{(2)}_{d} \,(E + m \tilde \phi) = (\Box + m^2) \, \bar C, \quad s^{(2)}_{d} E = \Box \bar C.
\end{eqnarray}
Under the above transformations, the Lagrangian density ${\cal L}_{(\bar B)}$ transforms as
\begin{eqnarray}
&&s^{(2)}_{ad} \,{\cal L}_{(\bar B)} = \partial_\mu \,\Bigl [ \,\bar {\cal B}\, \partial^\mu \,C 
+ m \, \varepsilon^{\mu\nu}\, \bigl (m \,A_\nu \,C - \phi \,\partial_\nu\, C \bigr )
- m\, \tilde \phi\, \partial^\mu\, C \Bigr ], \nonumber\\ 
&& s^{(2)}_d \,{\cal L}_{(B)} = \partial_\mu \,\Bigl [ \,\bar {\cal B}\, \partial^\mu \,\bar C 
+ m \, \varepsilon^{\mu\nu}\, \bigl ( m \,A_\nu \,\bar C - \phi\, \partial_\nu\, \bar C \bigr )
- m \,\tilde \phi \,\partial^\mu \,\bar C \Bigr ],
\end{eqnarray}
which establishes the symmetry invariance of the action integral of our present theory.

At this juncture, a few comments are in order. First and foremost, the 
nomenclature of the (anti-)co-BRST
symmetry transformations is very appropriate as we note that the total gauge-fixing term, owing its
fundamental origin to the co-exterior derivative, remains invariant under these transformations. Second,
the (anti-)co-BRST symmetry transformations are off-shell nilpotent (i.e. ($s^{(1,2)}_{(a)d} )^2 = 0$)
and they are absolutely anticommuting i.e. $s^{(1)}_d s^{(1)}_{ad} + s^{(1)}_{ad} s^{(1)}_d = 0, 
s^{(2)}_d s^{(2)}_{ad} + s^{(2)}_{ad} s^{(2)}_d = 0$) demonstrating  their fermionic and independent 
nature. Finally, the above absolute anticommutativity property 
ensures that only {\it one} of them could be identified with
the co-exterior derivative of differential geometry.

We wrap up this section with the remark that one could {\it also } talk about the symmetry properties of the Lagrangian densities
${\cal L}_{(B)}$ and ${\cal L}_{(\bar B)}$ under the transformations $s^{(2)}_{(a)d}$ and
$s^{(1)}_{(a)d}$, respectively. Taking into account the following nilpotent symmetry transformations
\begin{eqnarray}
&& s^{(1)}_d \bar B = 0, \qquad s^{(1)}_{ad}  \bar B = 0, \qquad
s^{(1)}_d \bar {\cal B} = 2 \, \Box\,\bar C, \qquad s^{(1)}_{ad} \bar {\cal B} = 2 \,\Box \,C,
\nonumber\\ 
&& s^{(2)}_d  B = 0, \qquad s^{(2)}_{ad}  B = 0, \qquad
s^{(2)}_d  {\cal B} = 2 \, \Box\,\bar C, \qquad s^{(2)}_{ad} {\cal B} = 2 \,\Box \,C,
\end{eqnarray}
in addition to the other transformations listed in (23) and (25), we observe the following 
interesting transformations for the Lagrangian densities:
\begin{eqnarray}
s^{(1)}_d \, {\cal L}_{(\bar B)} &=& \partial_\mu \,\Bigl [ {\cal B}\, \partial^\mu \,\bar C
+ m \,\varepsilon^{\mu\nu} \, \bigl (m \,A_\nu \,\bar C -  \phi\, \partial_\nu \,\bar C \bigr ) 
+ m \,\tilde \phi\, \partial^\mu \bar C \Bigr ] 
\nonumber\\ &-& \Bigl [ {\cal B} + \bar {\cal B} - 2 E \Bigr ] \,
(\Box + m^2 ) \bar C, \nonumber\\
s^{(1)}_{ad} \, {\cal L}_{(\bar B)} &=& \partial_\mu \,\Bigl [  {\cal B}\, \partial^\mu \, C
+ m \,\varepsilon^{\mu\nu} \, \bigl (m \,A_\nu \, C -  \phi\, \partial_\nu \, C \bigr ) 
+ m \,\tilde \phi\, \partial^\mu \, C \Bigr ] 
\nonumber\\ &-& \Bigl [ {\cal B} + \bar {\cal B} - 2 E \Bigr ] \,
(\Box + m^2 )  C, \nonumber\\
s^{(2)}_d \, {\cal L}_{( B)} &=& \partial_\mu \Bigl [  \bar {\cal B}\, \partial^\mu\, \bar C
+ m \,\varepsilon^{\mu\nu}\,  \bigl (m\, A_\nu \,\bar C +  \phi\, \partial_\nu \,\bar C \bigr ) 
- m \,\tilde \phi\, \partial^\mu \, \bar C \Bigr ]
\nonumber\\ &-& \Bigl [ {\cal B} + \bar {\cal B} - 2 E \Bigr ] \,
(\Box + m^2 ) \bar C, \nonumber\\
s^{(2)}_{ad} \, {\cal L}_{( B)} &=& \partial_\mu \Bigl [ \bar {\cal B}\, \partial^\mu\,  C
+ m \,\varepsilon^{\mu\nu}\,  \bigl (m\, A_\nu \, C +  \phi\, \partial_\nu \, C \bigr ) 
- m \,\tilde \phi\, \partial^\mu \,  C \Bigr ]
\nonumber\\ &-& \Bigl [ {\cal B} + \bar {\cal B} - 2 E \Bigr ] \,
(\Box + m^2 ) C, 
\end{eqnarray}
which demonstrate that 
we have symmetry invariance of the Lagrangian 
densities ${\cal L}_{(B)}$ and ${\cal L}_{(\bar B)}$
under the (anti-)co-BRST symmetry transformations $s^{(2)}_{(a)d}$ and
$s^{(1)}_{(a)d}$, too, provided we confine ourselves to a constrained hypersurface that is
described by the CF-type of field equations ${\cal B} + \bar {\cal B} - 2 E = 0$ in the 2D
Minkowskian spacetime manifold. This observation is exactly like the role played by the 
{\it original} 
CF-condition [17] in the context of 4D non-Abelian gauge theory discussed under 
the purview of BRST
formalism (see, e.g. [21,22] for details). However, we lay stress on the fact that the CF-type condition
of our present theory does 
{\it not} play any role in the proof of anticommutatvity property of
the (anti-)co-BRST symmetry transformations as is the case with the {\it original}
 CF-condition [17]
which {\it does} 
play a crucial role in the above proof for the (anti-)BRST symmetry transformations in the 
description of the 4D 1-form ($A^{(1)} = dx^\mu A_\mu$) non-Abelian gauge theory.

\section{Bosonic symmetries}

We have seen that there are four fermionic 
[$(s_{(a)b}^{(1,2)})^2 = 0,  (s_{(a)d}^{(1,2)})^2 = 0$] 
type symmetries for each of the Lagrangian densities
${\cal L}_{(B)}$ and ${\cal L}_{(\bar B)}$. It can be readily checked that, in their operator
form, these transformations obey the following relationships:
\begin{eqnarray}
&&\{s_b^{(1)} , s_{ab}^{(1)} \} = \{s_b^{(1)} , s_{ad}^{(1)} \} = \{s_d^{(1)} , s_{ad}^{(1)} \}
= \{s_d^{(1)} , s_{ab}^{(1)} \} = 0, \nonumber\\ 
&&\{s_b^{(2)} , s_{ab}^{(2)} \} = \{s_b^{(2)} , s_{ad}^{(2)} \} = \{s_d^{(2)} , s_{ad}^{(2)} \}
= \{s_d^{(2)} , s_{ab}^{(2)} \} = 0.
\end{eqnarray}
The following two independent anticommutators define the two bosonic symmetry transfromations
for the two equivalent Lagrangian densities. These are:
\begin{eqnarray}
\{s_b^{(1)}, s_d^{(1)} \} = s^{(1)}_w \equiv - \{s_{ab}^{(1)}, s_{ad}^{(1)} \}, \qquad
\{s_b^{(2)}, s_d^{(2)} \} = s^{(2)}_w \equiv - \{s_{ab}^{(2)}, s_{ad}^{(2)} \}. 
\end{eqnarray}
Thus, we note that there are unique bosonic symmetry transformations
for each of the Lagrangian  densities ${\cal L}_{(B)}$ and ${\cal L}_{(\bar B)}$ of our present theory.

The relevant fields of the above Lagrangian densities transform, under the infinitesimal
and continuous version of the bosonic transformations $s_w^{(1)}$, as
\begin{eqnarray}
&& s_w^{(1)} A_\mu = \partial_\mu {\cal B} + \varepsilon_{\mu\nu} \partial^\nu B, \qquad
s^{(1)}_w \phi = m\, {\cal B}, \qquad s^{(1)}_w \tilde \phi = m \, B, \nonumber\\
 && s^{(1)}_w (\partial \cdot A) = \Box\, {\cal B}, \qquad s^{(1)}_w\, E = - \Box B,
\qquad s^{(1)}_w \bigl [B, {\cal B}, C, \bar C \bigr ] = 0, \nonumber\\ 
&& s_w^{(2)} A_\mu = \partial_\mu \bar {\cal B} + \varepsilon_{\mu\nu} \partial^\nu \bar B, \qquad
s^{(2)}_w \phi = - m\, \bar {\cal B}, \qquad s^{(2)}_w \tilde \phi = - m \, \bar B, \nonumber\\
 && s^{(2)}_w (\partial \cdot A) = \Box\, \bar {\cal B}, \qquad s^{(2)}_w\, E = - \Box \bar B,
\qquad s^{(2)}_w \bigl [\bar B, \bar {\cal B}, C, \bar C \bigr ] = 0.  
\end{eqnarray}
One of the decisive features of these bosonic symmetry transformations  is the observation
that the ghost part of Lagrangian densities ${\cal L}_{(B)}$ and ${\cal L}_{(\bar B)}$
remains invariant. It can be readily checked that the above Lagrangian densities
transform to the total spacetime derivatives under the bosonic symmetry transformations.
These can be expressed mathematically as:  
\begin{eqnarray}
s_w^{(1)} \, {\cal L}_{(B)} &=& \partial_\mu \, \Bigl [ B \,\partial^\mu \,{\cal B} 
- {\cal B}\, \partial^\mu \,B
- m \,\varepsilon^{\mu\nu}\, \bigl (\phi \,\partial_\nu \,B + m\, A_\nu\, B \bigr ) 
- m \,\tilde \phi \,\partial^\mu \,B \Bigr ],
\nonumber\\
s_w^{(2)} \, {\cal L}_{(\bar B)} &=& \partial_\mu \, \Bigl [ \bar B\, \partial^\mu \,
\bar {\cal B} - \bar {\cal B} \,\partial^\mu \,\bar B
+ m \,\varepsilon^{\mu\nu} \,\bigl (\phi \,\partial_\nu \,\bar B - m \,A_\nu \,\bar B \bigr ) 
+ m \,\tilde \phi\, \partial^\mu\, \bar B \Bigr ],
\end{eqnarray}
ensuring the invariance of the action integrals for our current theory under consideration.

A few key points, at this stage, are as follows. First, the bosonic symmetry transformations
are unique because their two representations 
are equal modulo a sign factor. In other words, we
have: $s^{(1)}_w = \{ s_b^{(1)}, s_d^{(1)} \} \equiv - \{ s_{ab}^{(1)}, s_{ad}^{(1)} \}, \,
s^{(2)}_w = \{ s_b^{(2)}, s_d^{(2)} \} \equiv - \{ s_{ab}^{(2)}, s_{ad}^{(2)} \}$. Second,
the bosonic symmetry transformations are derived from the fundamental 
off-shell nilpotent (anti-)BRST
and (anti-)co-BRST symmetry transformations. Third, it is clear that the bosonic symmetry
transformations provide the physical realization of the Lapalacian operator of differential
geometry if one of the (anti-)BRST symmetry transformations is chosen to be the exterior
derivative {\it and} a single
 transformation from  the (anti-)co-BRST symmetry transformations is selected 
to represent the co-exterior derivative. Finally, we would like to state that the bosonic
transformations (31) are written modulo an overall $(-i)$ factor so that the conserved
Noether charge, corresponding to these transformations, could turn out to be real. This is due
to the fact that (identified with the Laplacian operator), this Noether 
charge should be Hermitian and it should produce the positive eigen values.

\section{ Ghost-scale and discrete symmetries}

We note that the  ghost part of the Lagrangian densities has the discrete symmetry 
transformations: $ C \to \mp i\, \bar C, \bar C \to \mp i\,  C$. Furthermore, there
are continuous scale transformations in the theory, under which, the relevant
fields of the theory transform as
\begin{eqnarray}
C \to e^{(+\,1\,) \Omega} \; C, \quad  \bar C \to e^{(-\,1\,) \Omega} \; C, \quad
\Psi \to e^{(\,0\,) \Omega}\, \Psi, \qquad \Psi \equiv A_\mu, B, \bar B, 
{\cal B}, \bar {\cal B}, \phi, \tilde \phi,
\end{eqnarray} 
which are called as the ghost-scale symmetry transformations because only the 
(anti-)ghost fields $(\bar C)C$, with ghost numbers $(\mp 1)$, transform and
rest of the fields of the theory, with zero ghost number, remain unchanged. 
The ghost numbers are reflected in the above scale transformations as the coefficients
of the parameter $\Omega$ in the exponentials. In the above,
the transformation scale parameter $\Omega$ is global (i.e. spacetime independent). The
infinitesimal version of these transformations, denoted by $s_g$,  are as follows:
\begin{eqnarray}
s_g C = + \, C, \qquad s_g \bar C = - \, \bar C, \qquad s_g \Psi = 0,
\end{eqnarray}
where 
we have chosen $\Omega = 1$ for the sake of brevity (for our further discussions). 
We have {\it not}
used any superscript for $s_g$ because the ghost part of the Lagrangian densities (15) and (18)
is the same and infinitesimal transformations (34) is true for both of them.

There are a couple of discrete symmetry transformations in our theory which, as we shall
see later, play very important roles in our whole discussion. In this context, let us recall 
the discrete transformations (8), under which, the Lagrangian densities (without the ghost part)
were found to be invariant. It can be checked that its generalized form
\begin{eqnarray}
&& A_\mu \to \pm i \varepsilon_{\mu\nu} A^\nu, \quad \phi \to \pm i \tilde \phi, 
\quad \tilde \phi \to \pm i  \phi, \quad E \to \mp i (\partial \cdot A), \quad
(\partial \cdot A) \to \mp i E, \nonumber\\
&& B \to \pm\, i\, {\cal B}, \;\; \bar B \to \pm\, i\, \bar {\cal B},
\; \;  {\cal  B} \to \pm\, i\, B, \; \;\bar {\cal  B} \to \pm\, i\, \bar B, \; \;
C \to \mp\, i\, \bar C, \;\; \bar C \to \mp\, i\, C,
\end{eqnarray}  
leaves the (anti-)BRST invariant Lagrangian densities (15) and (18) {\it absolutely} invariant.
We shall see that the above discrete symmetry transformations would provide the physical
realization of the Hodge duality operation of differential geometry (see, Sec. 7 below).

\section{Symmetries and cohomological operators}

The six (i.e. four fermionic and two bosonic)
continuous symmetry transformations of our present theory obey the following 
algebra in their operator form\footnote{We ignore here the superscripts $(1, 2)$ on the
continuous and infinitesimal
(anti-)BRST, (anti-)co-BRST and bosonic symmetry transformations because our statements,
in this section, are general in nature and true for both of them. In fact, our observations
are valid for any arbitrary model of Hodge theory.}: 
\begin{eqnarray}
&& s_b^2 = s_d^2 = s_{ab}^2 = s_{ad}^2 = 0, \qquad \{s_b, s_{ab} \} = 0, \qquad
\{s_d, s_{ad} \} = 0, \qquad \{s_b, s_{ad} \} = 0, \nonumber\\
&& \{s_d, s_{ab} \} = 0, \quad \{s_b, s_{d} \} = s_w = -  \{s_{ab}, s_{ad} \}, \quad
[s_w, s_r] = 0, \quad r = (b, ab, d, ad, g), \nonumber\\
&& [s_g, s_b ] = + s_b, \qquad [s_g, s_d ] = - s_d, \qquad  [s_g, s_{ad} ] = + s_{ad},  
\qquad [s_g, s_{ab}] = - s_{ab}.
\end{eqnarray}
The noteworthy points, at this stage, are as follows. First, we note that there 
are four fermionic and two bosonic type of continuous symmetries in our theory. Second,
the bosonic symmetry operator $s_w$ is the Casimir operator for the whole alegebra.
Finally, it is very important to note that
the ghost symmetry transformation $s_g$ has exactly similar kind of algebra with the
pair of symmetry operators $(s_b, s_{ad})$ and $(s_d, s_{ab})$.

A close look at the above algebra, satisfied by the continuous symmetry operators,
demonstrates that the structure of this algebra 
is very similar to the algebra of de Rham cohomological operators
$(d, \delta, \Delta)$ of differential geometry where, as we have seen earlier.
the operators $(\delta)d$ 
are the (co-)exterior derivatives and $\Delta = (d + \delta)^2$ is the Laplacian 
operator [23-25]. They obey the following well-known algebra
\begin{eqnarray}
d^2 = 0, \qquad \delta^2 = 0, \qquad \Delta = \{d, \delta \}, \qquad [\Delta, d] = 0, \qquad
[\Delta, \delta ] = 0.
\end{eqnarray}
Thus, we note that the Laplacian operator $\Delta$ is the Casimir operator for the whole
algebra and it is very similar to the bosonic symmetry transformation $s_w$ of the
algebra in (36). However, as we have seen earlier, there are two realizations of
$s_w$ in terms of the fermionic symmetry transformations: $s_w = \{ s_b, s_d \} =
- \{ s_{ab}, s_{ad} \}$. In contrast, we have only a single realization of $\Delta$
because $\Delta = \{d, \delta \}$ in differential geometry where $(\delta)d$ are nilpotent of order two (i.e. $d^2 = \delta^2 = 0$) just like the above fermionic symmetry operators. 
This shows that there is one-to-two mapping: $\Delta \Rightarrow \{s_b, s_d \} \equiv
- \{s_{ad}, s_{ab} \}$.

The above analysis leads us to make a guess that there
should be two realizations of $d$ and $\delta$ as well in the language of the symmetry transformations. It turns out that, this precisely is the case, in our 2D theory.
Thus, for our 2D model for the Hodge theory, the mapping is one-to-two from the cohomological operators to the symmetry operators, as 
\begin{eqnarray}
d \quad \Rightarrow \quad (s_b, s_{ad}), 
\qquad \delta \quad \Rightarrow \quad (s_d, s_{ab}), 
\qquad \Delta \quad \Rightarrow \quad s_w = \{ s_b, s_d \} =  -\, \{ s_{ab}, s_{ad} \}.
\end{eqnarray}
We wish to lay emphasis on the fact that it is the nature of the
commutation relations with $s_g$ that decides the grouping of the transformations
$(s_{b}, s_{ad})$ and $(s_d, s_{ab})$ in providing the realizations of the exterior
and co-exterior derivatives of  differential geometry.

We know that (co-)exterior derivatives $(\delta)d$ are connected by 
the relationship: $\delta = -\, *\, d \,*$ in the 
{\it even} dimensional spacetime manifold [23-25].
This relationship could be also realized in the language of symmetry transformations. 
For instance, it can be checked that the interplay between the continuous and discrete symmetry
transformations provide the realizations of the above relation (of differential geometry)
in the physical language of symmetries (of our present theory), as:
\begin{eqnarray}
s_{(a)d}\; \Psi \, = \,- \;*\;s_{(a)b}\; * \; \Psi, \qquad 
\qquad \Psi = A_\mu, \phi, \tilde \phi,
C, \bar C, B, \bar B, {\cal B}, \bar {\cal B}, E, (\partial \cdot A),
\end{eqnarray}
where $\Psi$ is the generic field of our present theory and $*$ operation is nothing
but the operation of the discrete symmetry transformations (35). The minus sign, present on the
r.h.s. of equation (39), is actually dictated by the two successive operations of the 
discrete symmetry transformations on any individual field. For instance, in our 2D theory,
it can be checked that the generic field $\Psi$ transforms as follows under the two successive
operations of the discrete symmetry transformations (35), namely;  
\begin{eqnarray}
*\; \bigl (\; *\; \Psi \; \bigr ) = -\; \Psi, \qquad 
\qquad \Psi = A_\mu, \phi, \tilde \phi,
C, \bar C, B, \bar B, {\cal B}, \bar {\cal B}, E, (\partial \cdot A).
\end{eqnarray}
Thus, we conclude that the relationship, quoted in equation (39), is correct as per the rules
and prescriptions laid down by the perfect duality invariant theory (see, e.g. [26]).

To sum up,
we have captured all the relevant {\it abstract} mathematical relations, obeyed by the 
de Rham cohomological operators of differential geometry, 
in the language of the discrete and continuous symmetry
transformations of our present theory which is {\it unique}, in some sense, because here the
{\it mass} and analogue of the gauge symmetries (and other symmetries) co-exist together
in a meaningful and complementary manner.

\section {Conclusions}

We have been able to establish, within the framework of BRST formalism, 
the 1D, 2D, 4D and 6D physically interesting  models to be 
the tractable physical examples of Hodge theory.
In this context, mention can be made of the 1D model of a rigid rotor [27], 2D free
(non-)Abelian gauge theories (without any interaction with matter fields) and the modified
version of 2D anomalous gauge theory [28], free 4D Abelian 2-form gauge theory
[5,6]  and free 6D
model of Abelian 3-form gauge theory [7,8]. Recently, we have shown the supersymmetric
quantum mechanical models to be an interesting set of examples for the Hodge theory [9-11].
However, we wish to re-emphasize that our present model of the modified
2D Proca theory is a very {\it special} 
example because, in this model, {\it mass}  and
various kinds of continuous and discrete symmetries co-exist {\it together} in a
meaningful manner.

One of the very important observations of our present investigation is the existence
of a couple of equivalent 
Lagrangian densities for the massive 2D Abelian 1-form gauge theory where
a set of (anti-)BRST invariant CF-type restrictions exist. Even though the latter do
not play any important role in the proof of anticommutatvity of the off-shell nilpotent
(i.e. $s_{(a)b}^2 = s_{(a)d}^2 = 0$) (anti-)BRST
and (anti-)co-BRST transformations of our present theory, they appear
when we discuss the symmetry properties of the Lagrangian densities {\it together}
under the (anti-)BRST and (anti-)co-BRST transformations. These are
completely novel observations in the context of the application of BRST formalism
to the description of a {\it massive} Abelian gauge theory. In our present investigation,
we have laid emphasis on these issues (in the main body of our text) in the language
of the off-shell nilpotent (anti-)BRST, (anti-)co-BRST, a unique bosonic and the ghost-scale
symmetries of the theory.

It is very interesting (and physically important) to 
point out that the kinetic term for the pseudo-scalar field 
$\tilde \phi$, in our theory, carries a negative sign. In the literature, such 
kinds of fields and particles have been discussed within the realm of quantum field 
theory and quantum mechanics (see, e.g. [29,30] for details). If the symmetries
are the guiding principles for a beautiful theory, it is essential to invoke a pseudo-scalar
field, with a negative kinetic term, in our theory so that we could have a set of perfect
discrete symmetry transformations
 [e.g. equation (35)] in our theory. We note that the equation of motion for
this field [i.e. $(\Box + m^2)\; \tilde \phi = 0$] shows that this intriguing field
carries a physical mass $m$. As a consequence, this field will provide a candidate for
the dark matter. Such kinds of {\it massless}
fields have also appeared in the context of 4D free Abelian
2-form gauge theory when we have proved it to provide a tractable model for the Hodge theory [7]. 

It is worthwhile to point out that the Stueckelberg real-scalar field, with positive
kinetic term, couples basically with the gauge-fixing ($\partial \cdot A)$ term as
is clear from a close look at the (anti-)BRST invariant Lagrangian densities (15)
and (18). On the contrary, the pseudo-scalar field, with negative kinetic term, 
couples with the electric field which happens to be a pseudo-scalar field in our
present 2D theory. We note that the coupling constant of these interaction terms
is nothing but the mass $m$ itself. This demonstrates that the fields ($\phi, \tilde \phi$)
can interact only via gravitational interaction. 
It would be very interesting to explore more details about these
fields in the physical four dimensions of spacetime which might give some clues
about the nature of the dark matter and its interaction with gauge fields. This aspect
of our observation would be
a key topic of research in our future endeavors.

During the past, we have exploited the theoretical power and potential of the
augmented version [31-33] of the usual superfield formalism [34-37] to derive the 
nilpotent (anti-)BRST and (anti-)co-BRST symmetry transformations for the gauge,
corresponding (anti-)ghost and matter fields of a given gauge/reparametrization
invariant theory. We are currently deeply involved with this aspect of
investigation in the context of the modified version of 2D Proca theory
and we have already obtained some novel results that are connected with the geometrical
interpretation of (and inter-relations between) the nilpotency and 
absolute anticommutativity
properties associated with the (anti-)BRST and (anti-)co-BRST  charges. We hope to report
about the key results of our work in the near future.

It would be very nice to extend our present analysis and ideas to 3D and 4D massive
gauge theories like Jackiew-Pi [38] and Freedman-Townsend [39] models of 
non-Abelian 1-form and 2-form theories. The 4D Abelian model
of the topologically massive gauge theory (with celebrated $B \wedge F$ term) is another
model where our ideas could be applied. The above models of
physical interest could be also proven, perhaps, to be the
physical examples of Hodge theory. These are the issues that are presently
 under investigation and
our results would be reported in our future publications [40].

\end{document}